\documentclass[showpacs,10pt,twocolumn,prb]{revtex4-1}
\usepackage{amsmath}
\usepackage{amssymb}
\usepackage{graphics}
\usepackage{epsfig}

\begin{document}

\title{New layered fluorosulfide SrFBiS$_{2}$}
\author{Hechang Lei, Kefeng Wang, Milinda Abeykoon, Emil S. Bozin, and C. Petrovic}
\affiliation{Condensed Matter Physics and Materials Science Department, Brookhaven National Laboratory, Upton, NY 11973, USA}
\date{\today}

\begin{abstract}
We have synthesized a new layered BiS$_{2}$-based compound SrFBiS$_{2}$.
This compound has similar structure to BiS$_{2}$. It is built up by
stacking up SrF layers and NaCl-type BiS$_{2}$ layers alternatively along
the c axis. Electric transport measurement indicates that SrFBiS$_{2}$ is a
semiconductor. Thermal transport measurement shows that SrFBiS$_{2}$ has a
small thermal conductivity and large Seebeck coefficient. First principle
calculations are in agreement with experimental results and show that %
SrFBiS$_{2}$ is very similar to LaOBiS$_{2}$ which becomes superconductor
with F doping. Therefore, SrFBiS$_{2}$ may be a parent compound of new
superconductors.
\end{abstract}

\maketitle

\section{Introduction}

Low-dimensional superconductors with layered structure have been extensively
studied and still attract much interest due to their exotic superconducting
properties and mechanism when compared to conventional BCS superconductors.
The examples include high $T_{c}$ cuprates,\cite{Bednorz} Sr$_{2}$RuO$_{4}$,%
\cite{Maeno} Na$_{x}$CoO$_{2}\cdot $H$_{2}$O,\cite{Takada} and iron-based
superconductors.\cite{Kamihara} The discovery of LnOFePn (Ln = rare earth
elements, Pn = P, As) in particular revitalises the study of layered
compounds with mixed anions, paving a way to materials with novel physical
properties. For example, Ln$_{2} $O$_{2}$TM$_{2}$OCh$_{2}$ (TM = transition
metals, Ch = S, Se) show strong electron-electron interactions and Mott
insulating state on the two dimensional (2D) frustrated antiferromagnetic
(AFM) checkerboard spin-lattice.\cite{Mayer}$^{-}$\cite{Free} Very recently,
bulk superconductivity was found in BiS$_{2}$-type layered compounds with
mixed anions: Bi$_{4}$O$_{4}$S$_{3}$ and Ln(O,F)BiS$_{2}$.\cite{Mizuguchi1}$%
^{-}$\cite{Demura} Experimental and theoretical studies indicate that these
materials exhibit multiband behaviors with dominant electron carriers
originating from the Bi 6$p_{x}$ and 6$p_{y}$ bands in the normal state.\cite%
{Usui}$^{-}$\cite{Tan} On the other hand, compounds with mixed anions
exhibit remarkable flexibility of structure. Different two-dimensional (2D)
building blocks, such as [LnO]$^{+}$, [AEF]$^{+}$ (AE = Ca, Sr, Ba), [Ti$%
_{2} $OPn$_{2}$]$^{2-}$, [FePn]$^{-}$, and [TM$_{2}$OCh$_{2}$]$^{2-}$, can
sometimes be integrated to form new materials.\cite{Kamihara}$^{,}$\cite%
{Matsuishi}$^{-}$\cite{Liu2} Individual building blocks often keep their
structural and electronic properties after being combined together.\cite%
{Kabbour}

In this work, we report the discovery of a new BiS$_{2}$-based layered
compound SrFBiS$_{2}$. It contains NaCl-type BiS$_{2}$ layer and shows
semiconducting behavior with relatively large thermopower. Theoretical
calculation indicates that this compound is very similar to LnOBiS$_{2}$.

\section{Experiment}

\subsection{Synthesis.}

SrFBiS$_{2}$ polycrystals were synthesized by a two-step solids state
reaction. First, Bi$_{2}$S$_{3}$ was prereacted by reacting Bi needles
(purity 99.99\%, Alfa Aesar) with sulfur flakes (purity 99.99\%, Aldrich) in
an evacuated quartz tube at 600 ${^{\circ }}C$ for 10 h. Then Bi$_{2}$S$%
_{3} $ was mixed with stoichiometric SrF$_{2}$ (purity 99\%, Alfa Aesar) and
SrS (purity 99.9\%, Alfa Aesar) and intimately ground together using an
agate pestle and mortar. The ground powder was pressed into 10 mm diameter pellets. We used a maximum pressure of 5 tons. The pressed pellet was loaded
in an alumina crucible and then sealed in quartz tubes with Ar under the
pressure of 0.15 atmosphere. The quartz tubes were heated up to 600 ${%
{}^{\circ }}C$ in 10 h and kept at 600 ${{}^{\circ }}C$ for another 10 h.

\subsection{Structure and Composition Analysis.}

Phase identity and purity were confirmed by powder X-ray diffraction carried
out by a Rigaku Miniflex X-ray machine with Cu K$_{\alpha }$ radiation ($%
\lambda $ = 1.5418 \AA ). Structural refinement of powder SrFBiS$_{2}$\
sample was carried out by using Rietica software.\cite{Hunter} Synchrotron
X-ray experiment was conducted at 300 K on X17A beamline of the National
Synchrotron Light Source (NSLS) at Brookhaven National Laboratory (BNL). The
setup utilized X-ray beam 0.5 mm $\times $ 0.5 mm in size and $\lambda $ =
0.1839 \AA\ ($E$ = 67.4959 keV), conditioned by two-axis focusing with
one-bounce sagittally-bent Laue crystal monochromator, and Perkin-Elmer
image plate detector mounted perpendicular to the primary beam path. Finely
pulverized sample packed in cylindrical polyimide capillary 1mm in diameter
was placed 204 mm away from the detector. Multiple scans were performed to a
total exposure time of 120 $s$. The 2D diffraction data were integrated and
converted to intensity versus 2$\theta $ using the software FIT2D.\cite%
{Hammersley} The intensity data were corrected and normalized and converted
to atomic pair distribution function (PDF), $G(r)$, using the program
PDFgetX2.\cite{Qiu} The average stoichiometry of SrFBiS$_{2}$ polycrystal was determined by examination of multiple points using an energy-dispersive x-ray spectroscopy (EDX) in a JEOL JSM-6500 scanning electron microscope.

\subsection{Electrical and Thermal Transport Measurements.}

The sample pellets were cut into rectangular bar and the surface is polished by sandpaper. Thin Pt wires were attached using silver epoxy for four probe resistivity measurements. Electrical and
thermal transport measurements were carried out in Quantum Design Physical
Property Measurement System (PPMS-9).

\subsection{Band Structure Calculations.}

First principle electronic structure calculation were performed using
experimental crystallographic parameters within the full-potential
linearized augmented plane wave (LAPW) method \cite{wien2k1} implemented in
WIEN2k package.\cite{wien2k2} The general gradient approximation (GGA) of
Perdew \textit{et al}.,\cite{gga} was used for exchange-correlation
potential. The LAPW sphere radius were set to 2.5 Bohr for all atoms, and
the converged basis corresponding to $R_{min}k_{max}=7$ with additional
local orbital were used where $R_{min}$ is the minimum LAPW sphere radius
and $k_{max}$ is the plane wave cutoff.

\section{Results and Discussion}

\subsection{Structure and Compostion.}

Fig. 1(a) shows the powder XRD pattern of SrFBiS$_{2}$ measured by Rigaku
Miniflex. Almost all of reflections can be indexed using the P4/nmm space
group. The unidentified peaks belong to the second phase of Bi$_{2}$S$_{3}$. Using two-phase Le Bail fitting, the refined lattice parameters of SrFBiS$_{2}$ are $a$ = 4.084(2)
\AA\ and $c$ = 13.798(2) \AA. When compared to LaOBiS$_{2}$, the a-axial lattice parameter is larger and the c-axial one is slightly smaller.\cite{Mizuguchi2} The PDF structural
analysis was carried out using the program PDFgui.\cite{Farrow} The SrFBiS$%
_{2}$ data are explained well within the model having P4/nmm symmetry with $%
a $ = 4.079(2) \AA\ and $c$ = 13.814(5) \AA  ($R_{wp}$ = 0.138, $\chi^{2}$=0.024). It is consistent with the
fitting results obtained from Miniflex. The final fit is shown in Fig. 1(b),
and the results are summarized in Table 1. In addition to the principal
phase, the sample is found to have $\sim $ 16(1) wt\% of Bi$_{2}$S$_{3}$
impurity with Pnma symmetry, which is also observed in Fig. 1(a). Structure
of SrFBiS$_{2}$ is similar to LaOBiS$_{2}$, which is built up by stacking
the rock-salt-type BiS$_{2}$ layer and fluorite-type SrF layer alternatively
along the c axis as shown Fig. 1(c). The EDX spectrum of polycrystal confirms the presence of Sr, F, Bi, and S. The average atomic ratios determined from EDX are Sr : F : Bi : S = 1.00(4) : 1.00(9) : 1.03(5) : 1.88(4) when setting the content of Sr as 1. It confirms the formula of obtained compound is SrFBiS$_{2}$.

\begin{figure}[tbp]
\centerline{\includegraphics[scale=0.6]{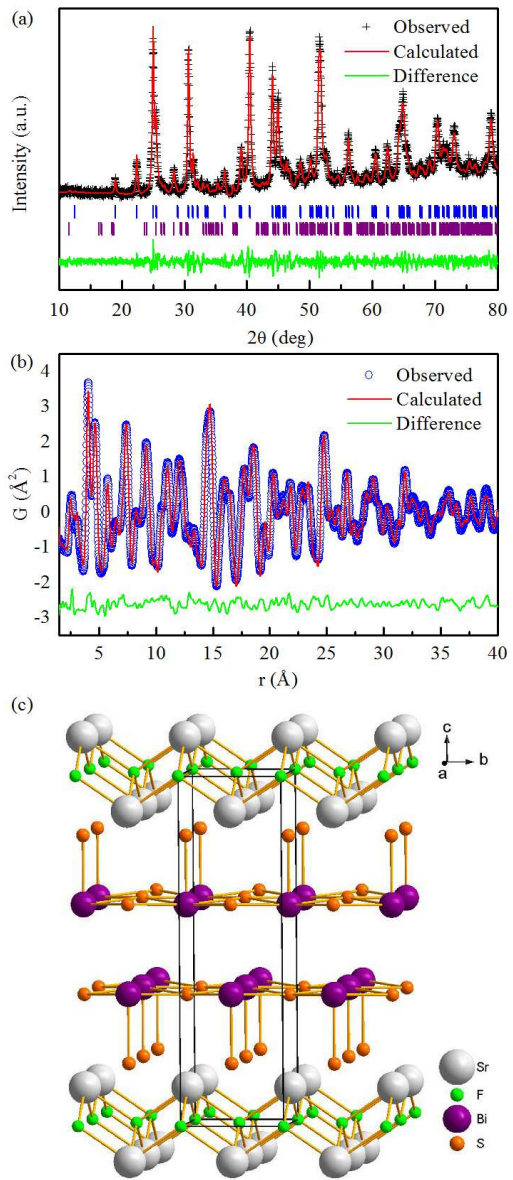}} \vspace*{-0.3cm}
\caption{(a) Powder XRD pattern of SrFBiS$_{2}$ and the fitted result using two-phase Le Bail fitting. Crosses are experimental data, red line is fitted spectra, green line is the difference between experimental data and fitted spectra, vertical lines are calculated Bragg-positions for SrFBiS$_{2}$ (upper) and Bi$_{2}$S$_{3}$ (lower), respectively. (b) Synchrotron PDF refinement of data taken at
room temperature. (c) Crystal structure of SrFBiS$_{2}$. The biggest white,
big purple, medium orange, and small green balls represent Sr, Bi, S, and F
ions, respectively.}
\end{figure}

\begin{table*}[tbp]\centering%
\caption{Crystallographic Data for SrFBiS$_{2}$ obtained from synchrotron
powder XRD.}%
\begin{tabular}{cccccc}
\hline\hline
\multicolumn{3}{c}{Chemical Formula} &  & \multicolumn{2}{c}{SrFBiS$_{2}$}
\\
\multicolumn{3}{c}{Formula Mass (g/mol)} &  & \multicolumn{2}{c}{379.73} \\
\multicolumn{3}{c}{Crystal System} &  & \multicolumn{2}{c}{Tetragonal} \\
\multicolumn{3}{c}{Space Group} &  & \multicolumn{2}{c}{P4/nmm (No. 129)} \\
\multicolumn{3}{c}{$a$ (\r{A})} &  & \multicolumn{2}{c}{4.079(2)} \\
\multicolumn{3}{c}{$c$ (\r{A})} &  & \multicolumn{2}{c}{13.814(5)} \\
\multicolumn{3}{c}{$V$ (\r{A}$^{3}$)} &  & \multicolumn{2}{c}{229.8(3)} \\
\multicolumn{3}{c}{Z} &  & \multicolumn{2}{c}{2} \\
\multicolumn{3}{c}{Density (g/cm$^{3}$)} &  & \multicolumn{2}{c}{5.51} \\
\hline
Atom & site & x & y & z & $U_{eq}$ (\r{A}$^{2}$)$^{a}$ \\
Sr & 2c & 1/4 & 1/4 & 0.1025(2) & 0.0069(4) \\
F & 2a & 3/4 & 1/4 & 0 & 0.033(2) \\
Bi & 2c & 1/4 & 1/4 & 0.6286(5) & 0.0183(3) \\
S1 & 2c & 1/4 & 1/4 & 0.379(3) & 0.060(2) \\
S2 & 2c & 1/4 & 1/4 & 0.811(2) & 0.019(1) \\ \hline\hline
\multicolumn{6}{c}{$^{a}$ $U_{eq}$ is defined as one-third of the
orthogonalized $U_{ij}$ tensor.}%
\end{tabular}%
\label{1}%
\end{table*}%

\begin{figure}[tbp]
\centerline{\includegraphics[scale=0.45]{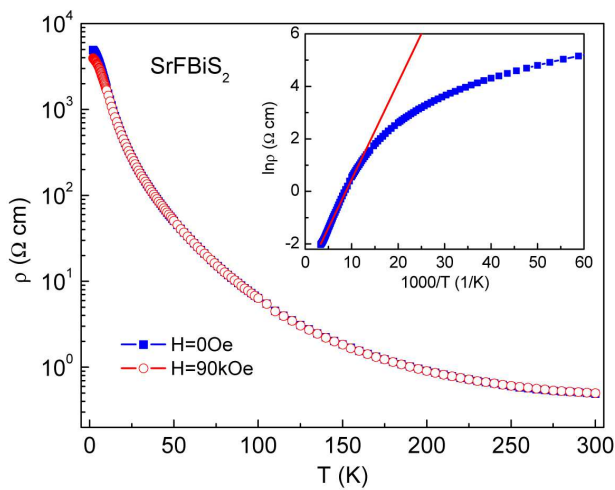}} \vspace*{-0.3cm}
\caption{Temperature dependence of the resistivity $\protect\rho (T)$ of the
SrFBiS$_{2}$ at $H$ = 0 (closed blue square)and 90 kOe(open red circle).
Inset shows the fitted result using thermal activation model for $\protect%
\rho (T)$ at zero field where the red line is the fitting curve.}
\end{figure}

\subsection{Electrical Properties.}

As shown in Fig. 2, the resistivity $\rho (T)$ of SrFBiS$_{2}$
polycrystalline shows a semiconducting behavior in the measured temperature
region (1.9-300 K). It should be noted that Bi$_{2}$S$_{3}$ polycrystal shows metallic
behavior becasue of sulfur deficiency.\cite{Chen} The impurity may have some
minor influence on the absolute value of resistivity, but the semiconducting
behavior should be intrinsic. Neglecting the grain boundary contribution,
the room-temperature resistivity $\rho (300$ K$)$ is about 0.5 $\Omega \cdot
$cm. Using the thermal activation model $\rho _{ab}(T)=\rho _{0}\exp
(E_{a}/k_{B}T)$ ($\rho _{0}$ is a prefactor, $E_{a}$ thermal activated
energy and $k_{B}$ the Boltzmann's constant) to fit the $\rho (T)$ at high
temperature (75 K - 300 K) (inset of Fig. 2), we obtain $E_{a}$ = 31.8(3)
meV. The semiconductor behavior is consistent with theoretical calculation
result shown below. On the other hand, theoretical calculations have
indicated that undoped LaOBiS$_{2}$ is also a semiconductor, which is partially consistent with the experimental result.\cite{Usui}$^{,}$\cite{Awana} Transport measurement indicates that LaOBiS$_{2}$ shows semiconducting behavior at $T <$ 200 K, but exhibit an upturn of resistivity at higher temperature. The origin of the upturn is unclear. Therefore, the replacement of LaO by SrF should not change
the band structure and thus physical properties too much, especially at low temperature, similar to the
relation between SrFFeAs and LaOFeAs.\cite{Han}$^{,}$\cite{Dong} The slight differences between LaOBiS$_{2}$ and SrFBiS$_{2}$, such as larger a-axial and smaller c-axial lattice parameters, could result in changing of physical properties at higher temperature. Note that the
semiconducting $\rho (T)$ in LaOBiS$_{2}$ and SrFBiS$_{2}$ are different
from those in parent compounds of iron pnictide superconductors. The latter
show metallic behaviors at high temperature and semiconducting-like upturn
in resistivity curve related to the spin density wave (SDW) transition.
There is no significant magnetoresistance in SrFBiS$_{2}$ up to 90 kOe
magnetic field.

\begin{figure}[tbp]
\centerline{\includegraphics[scale=0.3]{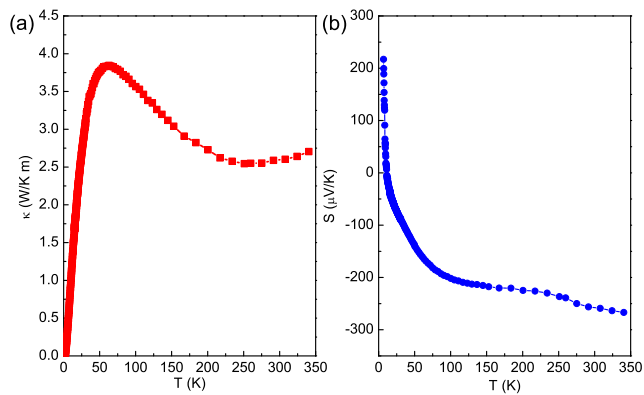}} \vspace*{-0.3cm}
\caption{Temperature dependence of (a) thermal conductivity and (b)
thermoelectric power for SrFBiS$_{2}$ under zero magnetic field within a
temperature range from 2 to 340 K.}
\end{figure}

\subsection{Thermal Transport Properties.}

The temperature dependences of the thermal conductivity $\kappa (T)$ and
thermoelectric power (TEP) $S(T)$ for SrFBiS$_{2}$ in zero field between 2
and 350 K are shown in Fig. 3. The electronic thermal conductivity $\kappa
_{e}(T)$ estimated from the Wiedemann-Franz law using a value for the Lorenz
number of 2.44$\times $10$^{-8}$ W $\Omega $/K$^{2}$ was less than 5$\times $%
10$^{-6}$ of $\kappa (T)$. Therefore, lattice thermal conductivity dominates
$\kappa _{L}(T)$ which exhibits a peak at around 60 K (Fig. 1(a)). The peak
in $\kappa (T)$ commonly arises since different phonon scattering processes
usually dominate in different temperature ranges. Umklapp scattering
dominates at high temperatures, while boundary and point-defect scattering
dominate at low and intermediate temperatures, respectively.\cite{Yang} On
the other hand, the $\kappa (T)$ of SrFBiS$_{2}$ shows similar behavior to Bi%
$_{4}$O$_{4}$S$_{3}$ but with different peak position and absolute value.%
\cite{Tan} For TEP $S(T)$ of SrFBiS$_{2}$, there is a reversal in sign at
about 11 K, i.e, hole-like carrier changes into electron-like carrier which
is dominant at room temperature. According to two band model, $%
S=|S_{h}|\sigma _{h}-|S_{e}|\sigma _{e}/(\sigma _{e}+\sigma _{h})$.\cite{Tan}
If we assume that $S_{h}$ and $S_{e}$ are temperature independent, it
suggests that electron and hole conductivities change dramatically with
temperature: at low temperature, $\sigma _{h}>\sigma _{e}$ whereas $\sigma
_{e}>\sigma _{h}$ above 11 K. Hole-like carrier may originate from defect
induced p-type doping. With increasing temperature, electron-like carrier
due to intrinsic band excitation increase significantly, finally leading to $%
\sigma _{e}>\sigma _{h}$ and a sign change in $S(T)$. Similar behavior was
observed in LaOZnP and p-type Si.\cite{Kayanuma,Seeger} Even though the $S(T)$ in SrFBiS$_{2}$ is significant and not much smaller than in classics
thermoelectric materials, \cite{Row} its low electrical conductivity makes
its figure of merit ZT (ZT = $\sigma $S$^{2}$T/$\kappa $) extremely small.

\begin{figure}[tbp]
\centerline{\includegraphics[scale=0.55]{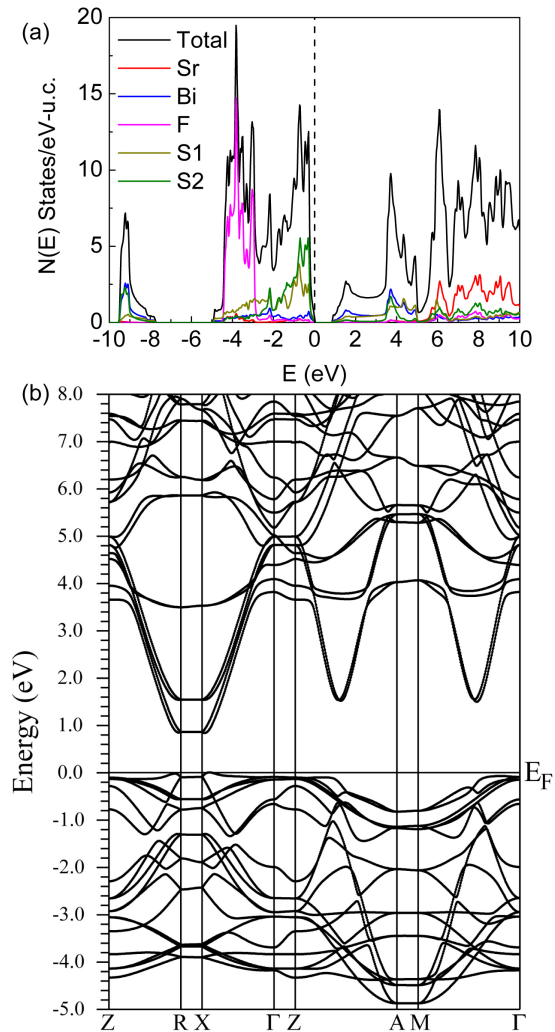}} \vspace*{-0.3cm}
\caption{(a) Total and atom resolved density of states and (b) band
structure of SrFBiS$_{2}$.}
\end{figure}

\subsection{Electronic Structure.}

First principle calculations (Fig. 4) confirm that SrFBiS$_{2}$ is a
semiconductor with a direct band gap of 0.8 eV located at $X$ point. This is
similar to LaOBiS$_{2}$ where the energy gap was found to be 0.82 eV.\cite%
{Wan} The calculation confirms the results of transport measurement. Similar
to LaOBiS$_{2}$,\cite{Usui}$^{,}$\cite{Wan} both S $3p$ and Bi $6p$ states
are located around the Fermi level (-2.0 to 2.0 eV) in SrFBiS$_{2}$. Thus
there is a strong hybridization between S $3p$ and Bi $6p$ states. The
absence of dispersion along $\Gamma -Z$ line suggests quasi two dimensional
character of the band structure in SrFBiS$_{2}$ (Fig. 4(b)). In LaOBiS$_{2}$, F doping results in metallic states and superconductivity at low
temperature. Main influence of F substitution is a carrier doping that
shifts the Fermi level and has only minor effect on the lowest conduction
band. Due to similarity between SrFBiS$_{2}$ and LaOBiS$_{2}$, new
superconductors could be obtained by chemical substitution.

\section{ Conclusion}

In summary, we report a discovery of a new layered fluorosulfide SrFBiS$_{2}$. It contains NaCl-type BiS$_{2}$ layer similar to Bi$_{4}$O$_{4} $S$_{3}$ and Ln(O,F)BiS$_{2}$ superconductors. SrFBiS$_{2}$\ polycrystals shows
semiconducting behavior between 2 K and 300 K. We observe rather small
thermal conductivity and large TEP with sign reversal at low temperature.
Theoretical calculation confirms the semiconducting behavior and indicates
similar DOS and band structure to undoped LaOBiS$_{2}$. Because of the
similarity between SrFBiS$_{2}$ and the parent compound of BiS$_{2}$-based
superconductors, it is of interest to investigate the doping effects on
physical properties of SrFBiS$_{2}$. It could pave a way to new members in
this emerging family of BiS$_{2}$-based superconductors.

\section{Acknowledgements}

We thank John Warren for help with SEM measurements. Work at Brookhaven is supported by the U.S. DOE under Contract No.
DE-AC02-98CH10886 and in part by the Center for Emergent Superconductivity,
an Energy Frontier Research Center funded by the U.S. DOE, Office for Basic
Energy Science (H. L. and C. P.). This work benefited from usage of X17A
beamline of the National Synchrotron Light Source at Brookhaven National
Laboratory. We gratefully acknowledge Zhong Zhong and Jonathan Hanson for
their help with the X17A experiment setup.


\begin{thebibliography}{99}

\bibitem{Bednorz} J. G. Bednorz and K. A. Muller, Z. Physik B \textbf{64},
189 (1986).

\bibitem{Maeno} Y. Maeno, H. Hashimoto, K. Yoshida, S. Nishizaki, T. Fujita,
J. G. Bednorz,\ and F. Lichtenberg, Nature \textbf{372}, 532 (1994).

\bibitem{Takada} K. Takada, H. Sakurai, E. Takayama-Muromachi, F. Izumi, R.
A. Dilanian, and T. Sasaki, Nature \textbf{422}, 53 (2003).

\bibitem{Kamihara} Y. Kamihara, T. Watanabe, M. Hirano, and H. Hosono, J.
Am. Chem. Soc. \textbf{130}, 3296 (2008).

\bibitem{Mayer} J. M. Mayer, L. F. Schneemeyer, T. Siegrist, J. V. Waszczak
and B. Van Dover, Angew. Chem., Int. Ed. Engl. \textbf{31}, 1645 (1992).

\bibitem{Wang} C. Wang, M. Q. Tan, C. M. Feng, Z. F. Ma, S. Jiang, Z. A. Xu,
G. H. Cao, K. Matsubayashi and Y. Uwatoko, J. Am. Chem. Soc. \textbf{132},
7069 (2010).

\bibitem{Zhu} J.-X. Zhu, R. Yu, H. Wang, L. L. Zhao, M. D. Jones, J. Dai, E.
Abrahams, E. Morosan, M. Fang and Q. Si, Phys. Rev. Lett. \textbf{104},
216405 (2010).

\bibitem{Ni} N. Ni, E. Climent-Pascual, S. Jia, Q. Huang and R. J. Cava,
Phys. Rev. B \textbf{82}, 214419 (2010).

\bibitem{Free} D. G. Free, N. D. Withers, P. J. Hickey and J. O. Evans,
Chem. Mater. \textbf{23}, 1625 (2011).

\bibitem{Mizuguchi1} Y. Mizuguchi, H. Fujihisa, Y. Gotoh, K. Suzuki, H.
Usui, K. Kuroki, S. Demura, Y. Takano, H. Izawa, O. Miura, Phys. Rev. B \textbf{86}, 220510(R) (2012).

\bibitem{Mizuguchi2} Y. Mizuguchi, S. Demura, K. Deguchi, Y. Takano, H.
Fujihisa, Y. Gotoh, H. Izawa, O. Miura, J. Phys. Soc. Jpn. \textbf{81}, 114725 (2012).

\bibitem{Demura} S. Demura, Y. Mizuguchi, K. Deguchi, H. Okazaki, H. Hara,
T. Watanabe, S. J. Denholme, M. Fujioka, T. Ozaki, H. Fujihisa, Y. Gotoh, O.
Miura, T. Yamaguchi, H. Takeya, and Y. Takano, arXiv 1207.5248.

\bibitem{Usui} H. Usui, K. Suzuki, K. Kuroki, Phys. Rev. B \textbf{86}, 220501(R) (2012).

\bibitem{Li} S. Li, H. Yang, J. Tao, X. Ding, and H.-H. Wen, arXiv 1207.4955.

\bibitem{Singh} S. K. Singh, A. Kumar, B. Gahtori, Shruti, G. Sharma, S.
Patnaik, and V. P. S. Awana, J. Am. Chem. Soc \textbf{134}, 16504 (2012).

\bibitem{Tan} S. G. Tan, L. J. Li, Y. Liu, P. Tong, B. C. Zhao, W. J. Lu, Y.
P. Sun, Physica C \textbf{483}, 94 (2012).

\bibitem{Matsuishi} S. Matsuishi, Y. Inoue, T. Nomura, H. Yanagi, M. Hirano,
and H. Hosono, J. Am. Chem. Soc. \textbf{130}, 14428 (2008).

\bibitem{Han} F. Han, X. Y. Zhu, G. Mu, P. Cheng, and H.-H. Wen, Phys. Rev.
B \textbf{78}, 180503 (2008).

\bibitem{Kabbour} H. Kabbour, L. Cario, and F. Boucher, J. Mater. Chem.
\textbf{15}, 3525 (2005).

\bibitem{Liu} R. H. Liu, J. S. Zhang, P. Cheng, X. G. Luo, J. J. Ying, Y. J.
Yan, M. Zhang, A. F. Wang, Z. J. Xiang, G. J. Ye and X. H. Chen, Phys. Rev.
B \textbf{83}, 174450 (2011).

\bibitem{Liu2} R. H. Liu, Y. A. Song, Q. J. Li, J. J. Ying, Y. J. Yan, Y.
He, and X. H. Chen, Chem. Mater. \textbf{22}, 1503 (2010).

\bibitem{Hunter} Hunter B. (1998) "Rietica - A visual Rietveld program",
International Union of Crystallography Commission on Powder Diffraction
Newsletter No. 20, (Summer) http://www.rietica.org

\bibitem{Hammersley} A. P. Hammersley, S.O. Svenson, M. Hanfland, and D.
Hauserman, High Press. Res. \textbf{14}, 235 (1996).

\bibitem{Qiu} X. Qiu, J. W. Thompson, and S. J. L. Billinge, J. Appl.
Crystallogr. \textbf{37}, 678 (2004).

\bibitem{wien2k1} M. Weinert, E. Wimmer, and A. J. Freeman, Phys. Rev. B
\textbf{26}, 4571 (1982).

\bibitem{wien2k2} P. Blaha, K. Schwarz, G. K. H. Madsen, D. Kvasnicka and J.
Luitz, WIEN2k, An Augmented Plane Wave + Local Orbitals Program for
Calculating Crystal Properties (Karlheinz Schwarz, Techn. Universitat Wien,
Austria), 2001. ISBN 3-9501031-1-2

\bibitem{gga} J. P. Perdew, K. Burke and M. Ernzerhof, Phys. Rev. Lett.
\textbf{77}, 3865 (1996).

\bibitem{Farrow} C. L. Farrow, P. Juhas, J. W. Liu, D. Bryndin, E. S. Bozin,
J. Bloch, Th. Proffen, and S. J. L. Billinge, J. Phys.: Condens. Mater.
\textbf{19}, 335219 (2007).

\bibitem{Chen} B. Chen, C. Uher, L. Iordanidis, and M. G. Kanatzidis, Chem.
Mater. \textbf{9}, 1655 (1997).

\bibitem{Awana} V. P. S. Awana, A. Kumar, R. Jha, S. Kumar, J. Kumar, and A.
Pal, arXiv 1207.6845.

\bibitem{Dong} J. Dong, H. J. Zhang, G. Xu, Z. Li, G. Li, W. Z. Hu, D. Wu,
G. F. Chen, X. Dai, J. L. Luo, Z. Fang, and N. L. Wang, EPL \textbf{83},
27006 (2008).

\bibitem{Yang} J. Yang, D. T. Morelli, G. P. Meisner, W. Chen, J. S. Dyck,
and C. Uher, Phys. Rev. B \textbf{65}, 094115 (2002).

\bibitem{Seeger} K. Seeger, in Semiconductor Physics: An Introduction,
Springer-Verlag, Berlin (2004).

\bibitem{Kayanuma} K. Kayanuma, H. Hiramatsu, M. Hirano, R. Kawamura, H.
Yanagi, T. Kamiya, and H. Hosono, Phys. Rev. B \textbf{76}, 195325 (2007).

\bibitem{Row} D. M. Rowe, in Thermoelectrics Handbook: Macro to Nano, Taylor
\& Francis, London (2006).

\bibitem{Wan} X. Wan, H.-C. Ding, S. Y. Savrasov, and C.-G. Duan, arXiv
1208.1807.
\end{thebibliography}
\end{document}